\documentclass[
  aps,
  pre,
  twocolumn,
  superscriptaddress,
  groupedaddress
]{revtex4}

\usepackage{amsmath}    
\usepackage{graphicx}   
\usepackage{verbatim}   
\usepackage[usenames]{color}  
\usepackage{subfigure}  
\usepackage{hyperref}   
\usepackage{amssymb}
\usepackage{epstopdf}
\usepackage{float}

\begin{document}

\title{Effects of RNA branching on the electrostatic stabilization of viruses}
 \author{Gonca Erdemci-Tandogan}
 \author{Jef Wagner}
 \affiliation{Department of Physics and Astronomy,
   University of California, Riverside, California 92521, USA}
 \author{Paul van der Schoot}
 \affiliation{Group Theory of Polymers and Soft Matter, Eindhoven University of Technology, P.O. Box 513, 5600 MB Eindhoven,
   The Netherlands}
 \affiliation{ Institute for Theoretical Physics,
   Utrecht University,
   Leuvenlaan 4, 3584 CE Utrecht, The Netherlands}
 \author{Rudolf Podgornik}
 \affiliation{Department of Physics, University of Massachusetts, Amherst, MA 01003}
 \affiliation{Department of Theoretical Physics, J. Stefan Institute, SI-1000 Ljubljana, Slovenia}
 \affiliation{Department of Physics, University of Ljubljana, SI-1000 Ljubljana, Slovenia}
 \author{Roya Zandi}
 \affiliation{Department of Physics and Astronomy,
   University of California, Riverside, California 92521, USA}

\begin{abstract}
Many single-stranded (ss) RNA viruses self assemble from capsid protein subunits and the nucleic acid to form an infectious virion. It is believed that the electrostatic interactions between the negatively charged RNA and the positively charged viral capsid proteins drive the encapsidation, although there is growing evidence that the sequence of the viral RNA also plays a role in packaging. In particular the sequence will determine the possible secondary structures that the ssRNA will take in solution. In this work, we use a mean field theory to investigate how the secondary structure of the RNA combined with electrostatic interactions affects the efficiency of assembly and stability of the assembled virions. We show that the secondary structure of RNA may result in negative osmotic pressures while a linear polymer causes positive osmotic pressures for the same conditions. This may suggest that the branched structure makes the RNA more effectively packaged and the virion more stable.
\end{abstract}

\maketitle

%%%%%%%INTRODUCTION%%%%%%
\section{introduction}

Many single-stranded (ss) RNA viruses package their genome concurrently with the self-assembly of the whole capsid in such a way, that small protein subunits spontaneously assemble around the nucleic acid to built a complete protein shell (capsid) \cite{Rossmann}. In the prevailing paradigm this assembly is predominantly driven by generic, nucleotide sequence independent, electrostatic interactions \cite{Siberpccp} between the negative charges on the RNA phosphate backbone and the positive charges on the virus capsid proteins (CP) \cite{Bogdan, Vanderschoot2007, Anze2, Zlotnick, Hsiang-Ku,Kusters2015}. Recent experiments have indeed abundantly verified the importance of the ``charge-matching hypothesis", based on the preponderance of electrostatic interactions  between the capsid proteins and the RNA for proper genome packaging \cite{Garmann2014}.  

However, besides the importance of electrostatics, packaging experiments suggest that there must  exist a correlation between the specific details of the nucleic acid structure and the efficient virus assembly \cite{Comas,Yoffe2008,Li-tai,Venky2016}. In a beautifully designed experiment Comas-Garcia et al. \cite{Comas} have set the viral RNA1 of Brome Mosaic Virus (BMV) and the RNA of Cowpea Chlorotic Mottle Virus (CCMV) to compete against each other for capsid proteins belonging to CCMV exclusively. Although both RNAs are of similar length, BMV RNA was shown to outcompete the CCMV RNA, therefore  suggesting that electrostatics alone is not enough for efficient genome encapsidation and that further structural details of RNA, apart from its generic charge, could play a role in the genome encapsidation \cite{Comas,Garmann2015}. 

Even further away from the presumed non-specificity of the genome - CP interactions are indications, from both {\sl in vitro} and {\sl in vivo} studies, that the capsid self-assembly is achieved via a directed capsid assembly mediated by the highly specific, non-electrostatic interactions between sections of RNA and capsid proteins; these sections of RNA are thought to contain {\sl packaging signals} and are repeated along the genome according to the symmetry of the capsid \cite{Stockley}. Contrary to the generic electrostatic charge matching, the essence of the Òpackaging signal hypothesisÓ is thus that the viral genomes have local secondary or tertiary structures with high CP affinity, serving as heterogeneous nucleation sites for the formation of capsids \cite{Patel,Roya2006}. Quite interestingly, in a recent experiment on Satellite Tobacco Mosaic Virus (STMV), Sivanandam {\it et al.} find that reducing the number of charges on the N-terminal section of capsid proteins through mutations results into the encapsidation of shorter RNAs than the wild type ones. However, unexpectedly a single mutation in one specific location along the N-terminal completely stops the self-assembly \cite{Venky2016}. Investigating the nature of how and which structural details of RNA could be important for virus assembly is thus urgently required to ascertain on which point along the axis of ``charge-matching" to ``packaging signals" hypotheses the viruses actually drive and regulate their assembly. 

Viral RNAs are found to be compact and highly branched \cite{GarmannBranched} due to the base-pairing between the nucleotides, engendering compactification and folding of the molecule. Indeed, it appears that the compactness of the ssRNA wild-type viral genomes is one of the principal characteristics of their nucleotide sequence, setting them distinctly apart from randomized sequences \cite{Yoffe2008,Gonca2016}, and that the physical compactness of the viral genome can be regarded as a primary factor among evolutionary constraints \cite{LucaRudi2015}.

While theoretical arguments suggest that the details of the RNA structure are important for its efficient packaging in the small volume of the virus capsid \cite{vanderschoot2009,Paul:13a,Hagan,RNAtopology14,Ben-Shaul2015,Venky2016}, it remains overall poorly understood how the RNA sequence chemical composition together with its length affect the compactification and the packaging efficiency. Based on simple scaling arguments, it has been shown that genome secondary structures, or more specifically branching, lower the free energy of RNA encapsidation \cite{vanderschoot2009,Paul:13a}.  As far as the length of RNA is concerned, there is a clear correlation with the number of positive charges on the virus coat proteins, structurally due to their extended N-tails, for many ssRNA viruses \cite{Shklovskii,Belyi, Hagan,Siber-nonspecific,Paul:13a}. This correlation ratio is $\sim 1.6$ for many wild type viruses \cite{Belyi}, implying that the number of negative charges on the RNA is in fact larger than the number of positive charges on the protein motifs, making these viruses {\sl overcharged}. 

Furthermore, when virus coat proteins encapsidate a linear polymer, {\textit {e.g.}}, poly(styrene sulfonate) (PSS), two different results are obtained: both highly overcharged (correlation ratio $\sim 9$ \cite{Chuck2008}) and undercharged (correlation ratio between 0.45 and 0.6 \cite{Cadena2011}) virus-like particles (VLP). The overcharging phenomenon has been discussed in many theoretical papers with different conclusions dependening mostly on the details of the model under consideration \cite{Belyi,Shklovskii, Vanderschoot, Ting, Siber-nonspecific, Chuck2008,Cadena2011,Zandi2016}. What one would hope for is that the important characteristics of the RNA genome packaging would robustly depend on some well defined characteristics of the genome, a hypothesis recently proposed in our work \cite{RNAtopology14}, where we showed that the secondary structure of RNA, as quantified by its branchiness, coupled to electrostatic interactions enhances the genome encapsidation capacity and could robustly explain the overcharging actually observed in virions.

While understanding the detailed role of electrostatics and structure of RNA on the self-assembly is the focus of what follows, we also aim additionally to understand what controls the virions or VLP stability or what the main factors are that enhance this stability before the disassembly of the capsid. Viruses seem to release their genome during the disassembly \cite{Roos2007}, which would imply that the genome not just leaves, but is in fact actively pushed from the capsid - a scenario that has been shown as specifically valid for bacteriophages, where the repulsive DNA-DNA interactions act like a coiled osmotic spring ejecting the genome. The corresponding osmotic pressure is in fact quite large and positive, surpassing even 50 atm, and stemming mostly from the combination of electrostatic and hydration interactions that are dominant in the range of DNA densities relevant for bacteriophage packing \cite{Siberpccp}. 

Contrary to DNA in bacteriophages, the osmotic pressure in ssRNA viruses is not easy to measure directly and in the absence of experiments one thus has to rely on theoretical estimates. {There have been several theoretical studies that investigate the osmotic pressure of ssRNA viruses \cite{Bruinsma,Vanderschoot,Siber-nonspecific,Javidpour,Andoh2014}.} Siber and Podgornik showed that the filled ssRNA virions exhibit a small residual negative osmotic pressure, which depends strongly on the amount of capsid charges and can be turned positive with relatively higher capsid charge \cite{Siber-nonspecific}. In addition, Javidpour et al. studied the effects of multivalent ions, which can fundamentally change the nature of electrostatic interactions \cite{Perspective}, on the osmotic pressure and the stability of the virus like empty shells, showing that the multivalent ions can turn a positive electrostatic osmotic pressure into a negative one \cite{Javidpour}. Furthermore, recent all atom molecular dynamics simulations showed that the osmotic pressure inside an empty Poliovirus capsid is negative,  suggesting that the mechanism might be connected with excess charges on the capsid that prevent the solution ion to exchange with the capsid \cite{Andoh2014}, a scenario at odds with what we know about the permeability of capsids. While there have thus been several lines of investigation regarding the nature and specifically the sign of the capsid osmotic pressure, there exist no studies taking into account the role of the secondary structure of RNA in the osmotic pressure of ssRNA viruses or virus like particles, another aspect that we elucidate further below.

In this paper, we extend our previous analysis and investigate how the secondary structure of the RNA affects the osmotic pressure of ssRNA viruses and what are the repercussions for stability of the virions. We show that the secondary structure of RNA may indeed result in negative osmotic pressures at conditions where a linear polymer would exhibit positive osmotic pressures. This may suggest that having a branched structure makes not only RNA more effectively packaged but also makes a virion more stable.  The paper is organized as follows. In the next section, we introduce the model and the fundamentals of the theory together with the basic quantities that we will calculate. In Sec. \ref{results}, we present the results for osmotic pressure as well as the effect of RNA branching on the free energy minimum, defining the optimum length of RNA, the optimum number of branched points and the optimum charge ratios of the system, together with the corresponding ion  concentration and RNA density profiles. Section \ref{discussion} discusses effects of different models, boundary conditions and different parameterizations that might correspond to different types of viruses. Finally, we summarize our findings. In the appendix, we derive in detail the model free energy of the encapsidation.

%%%%%%%MODEL%%%%%%%%
\section{Model}\label{model}
To elucidate the role of genome in the assembly of spherical RNA viruses, we model RNA as a generic, negatively charged, flexible branched polyelectrolyte that interacts with positive charges residing on the inner surface of the capsid.  More specifically,  we consider only the case of {\sl annealed branched polymers} because the strength of RNA base-pairing is relatively weak and may easily be affected by the interaction with the positive inner surface charges of the shell during encapsidation.  For simplicity, we model the capsid as a thin sphere and assume that the charges are not localized but smeared out uniformly on the inner surface of the sphere. We note that while a thin shell is a good approximation for the capsid of some viruses like Dengue and yellow fever \cite{Anze}, the capsid proteins of some other viruses contain N-terminal tails which are highly positively charged and point into the capsid cavity in a brush-like fashion \cite{Shklovskii}. 

The mean-field free energy functional of a polyelectrolyte chain confined within a charged shell in a univalent salt solution, under the ground state approximation, can be written as 
\begin{multline} \label{free_energy}
  \beta F = \!\!\int\!\! {\mathrm{d}^3}{{{r}}}\Big[
    \tfrac{a^2}{6} |{\nabla\Psi({\bf{r}})}|^2
    +W\big[\Psi({\bf{r}}) \big]\\
    -\tfrac{\beta^2 e^2}{8 \pi \lambda_B} |{\nabla\Phi({\bf{r}})}|^2
    -2\mu\cosh\big[\beta e \Phi({\bf{r}})\big]
    + \beta \tau \Phi({\bf{r}})\Psi^2({\bf{r}})
    \Big]\\
  + \int\!\! {\mathrm{d}^2}r \Big[ \beta \sigma \, \Phi({\bf{r}}) \Big].
\end{multline}
Here $\beta$ denotes the inverse of the thermal energy $k_BT$, $a$ the statistical step (Kuhn) length of the polymer,  $\tau$ the linear charge density of the polymer, $\sigma$ the surface charge density of the shell, $\Psi({\bf{r}})$ the monomer density field at position $\mathbf{r}$, and $\Phi({\bf{r}})$ the mean electrostatic potential. The parameter $\mu$ is the fugacity of the monovalent salt ions corresponding to the concentration of salt ions in the bulk. $\lambda_B={e^2 \beta}/{4 \pi \epsilon \epsilon_0}$, is the Bjerrum length, a measure of the dielectric constant ($\epsilon$) of the solvent and is about $0.7$ nm for water at room temperature. 

The first term of Eq.~\eqref{free_energy} is the entropic cost of non-uniform polymer density and the last two lines of Eq.~\eqref{free_energy} correspond to the electrostatic interactions between the polymer, the shell and the salt ions on the level of the Poisson-Boltzmann theory \cite{Siber-nonspecific}. The standard form of this free energy can be found in references \cite{Borukhov,Siber-nonspecific}. For completeness we also provide a step by step derivation of Eq.~\eqref{free_energy} for a linear polymer in the appendix.

The self-interaction term $W[\Psi]$  in Eq.~\eqref{free_energy} is associated with the self repulsion of the polyelectrolyte and the energy of an annealed branched polymer \cite{Lubensky,Nguyen-Bruinsma,Lee-Nguyen,Elleuch}, 
\begin{align} \label{W_branched}
  W[\Psi]&=\frac{1}{2}\upsilon \Psi^4 -\frac{1}{\sqrt{a^3}}(f_e\Psi+\frac{a^3}{6} f_b \Psi^3),
\end{align}
where $\upsilon$ is the excluded volume term and $f_e$ and $f_b$ are the fugacities of the end- and branch-points of the annealed polymer, respectively. A detailed derivation of Eq.~\ref{W_branched} is given in \cite{adsorption2015}. In this model, the stem-loop or hair-pin configurations of RNA are counted as the end points. The number of end- and branch-points $N_e$ and $N_b$ of the polymer are related to the fugacities $f_e$ and $f_b$ in a standard way by 
\begin{align}\label{NeNb}
N_e =- \beta f_e \frac{\partial{F}}{\partial{f_e}} \qquad {\rm and} \qquad N_b =- \beta f_b \frac{\partial{F}}{\partial{f_b}}.
\end{align}
We have two additional constraints in the problem. First, the total number of monomers inside the capsid is fixed \cite{Hone},
\begin{equation}\label{constraint}
  N= \int {\mathrm{d}^3}{r} \; \Psi^2 ({\bf{r}}),	
\end{equation}
a constraint that we enforce by introducing a Lagrange multiplier, $E$, when minimizing the free energy. Second, the number of the end points depends on the number of branched points so that
\begin{equation}\label{branch_constraint}
  N_e = N_b+2,
\end{equation}
since we consider only a single polymer with no closed loops. Thus, $f_e$ is not a free parameter. For our calculations, we change $f_b$ and find $f_e$ through  Eqs.  ~\eqref{NeNb} and ~\eqref{branch_constraint}. The polymer is linear if $f_b=0$, and the number of branched points increases with $f_b$. 

By varying the free energy functional with respect to fields $\Psi({\bf r})$ and $\Phi({\bf r})$, we obtain a coupled set of non-linear differential equations coupling the monomer density with the electrostatic potential in the interior of the capsid, and the usual Poisson-Boltzmann equation for the exterior of the capsid.   The monomer density field in fact satisfies the modified Edwards equation
\begin{equation} 
    \frac{a^2}{6} \nabla^2 \Psi (\mathbf{r}) 
    =-E {\Psi} (\mathbf{r}) + \beta \tau \Phi_{in}(\mathbf{r}) \Psi(\mathbf{r})+
    \frac{1}{2} \frac{\partial W}{\partial \Psi}, 
    \label{euler_a}
 \end{equation}
while the electrostatic potential satisfies the modified Poisson-Boltzmann equation in the interior of the capsids 
\begin{equation} 
    \nabla^2 \Phi_{in} (\mathbf{r})  
    = \tfrac{1}{\lambda_D ^{2} \beta e}
    \sinh \big [ \beta e \Phi_{in}(\mathbf{r}) \big ]
    -  \tfrac{\tau}{{2 \lambda_D ^{2}} \mu \beta e^2}
    {\Psi}^2 (\mathbf{r}),
    \label{euler_b}
\end{equation}
and the standard Poisson-Boltzmann equation in the exterior
\begin{equation}     
    \nabla^2 \Phi_{out} (\mathbf{r})  
    = \tfrac{1}{\lambda_D ^{2} \beta e}
    \sinh \big [ \beta e \Phi_{out}(\mathbf{r}) \big ],
    \label{euler_c}
\end{equation}
where $\lambda_D=1/\sqrt{8\pi \lambda_B \mu}$ is the Debye screening length.  The boundary condition (BC) for the electrostatic potential is obtained by minimizing the free energy, ${\hat n\cdot}{\nabla \Phi_{in}}-{\hat n\cdot}{\nabla \Phi_{out}} = {{4\pi\lambda_B}}{\sigma}/\beta e^2$ assuming the surface charge density $\sigma$ is fixed. The concentration of the polymer outside of the capsid is assumed to be zero.  The BC for the inside monomer density field $\Psi$ is of Neumann type (${\hat n\cdot}{\nabla \Psi}|_s=0$), that can be obtained from the energy minimization \cite{Hone}. However, due to the short-ranged self-repulsions of the polymer, Dirichlet type BC ($\Psi|_s=0$) might be preferable so that the polymer density goes to zero on the surface of the capsid. In our calculations we use both types of BCs and find that our conclusions do not depend on their detailed nature so that our conclusions are robust. We start with the Neumann BC but discuss the impact of the Dirichlet BC later in Sec. \ref{discussion}.

Using Eq.~\eqref{free_energy}, we can also obtain the osmotic pressure due to the genome encapsidation, {\it i.e.}, the force exerted on the virus capsid by the genome per unit surface area, defined as
 \begin{equation} \label{eq:op}
	P(N)=- \bigg{(} \frac{\partial F}{\partial V}{\bigg|_{Q_c,N}} - \frac{\partial F}{\partial V}{\bigg|_{Q_c,N=0}} \bigg{)},
\end{equation}
where $V$ is the volume of the capsid and we subtracted the part of the osmotic pressure for the empty capsid. {In the calculation of the pressure, we keep the total number of monomers $N$ and the total number of charges on the capsid $Q_c=4 \pi b^2 \sigma$ constant with $b$ the radius of the capsid.}

%%%%%%%%%%RESULTS%%%%%%%%
\section{Results}\label{results}

\begin{figure}
  \includegraphics[width=8.6cm]{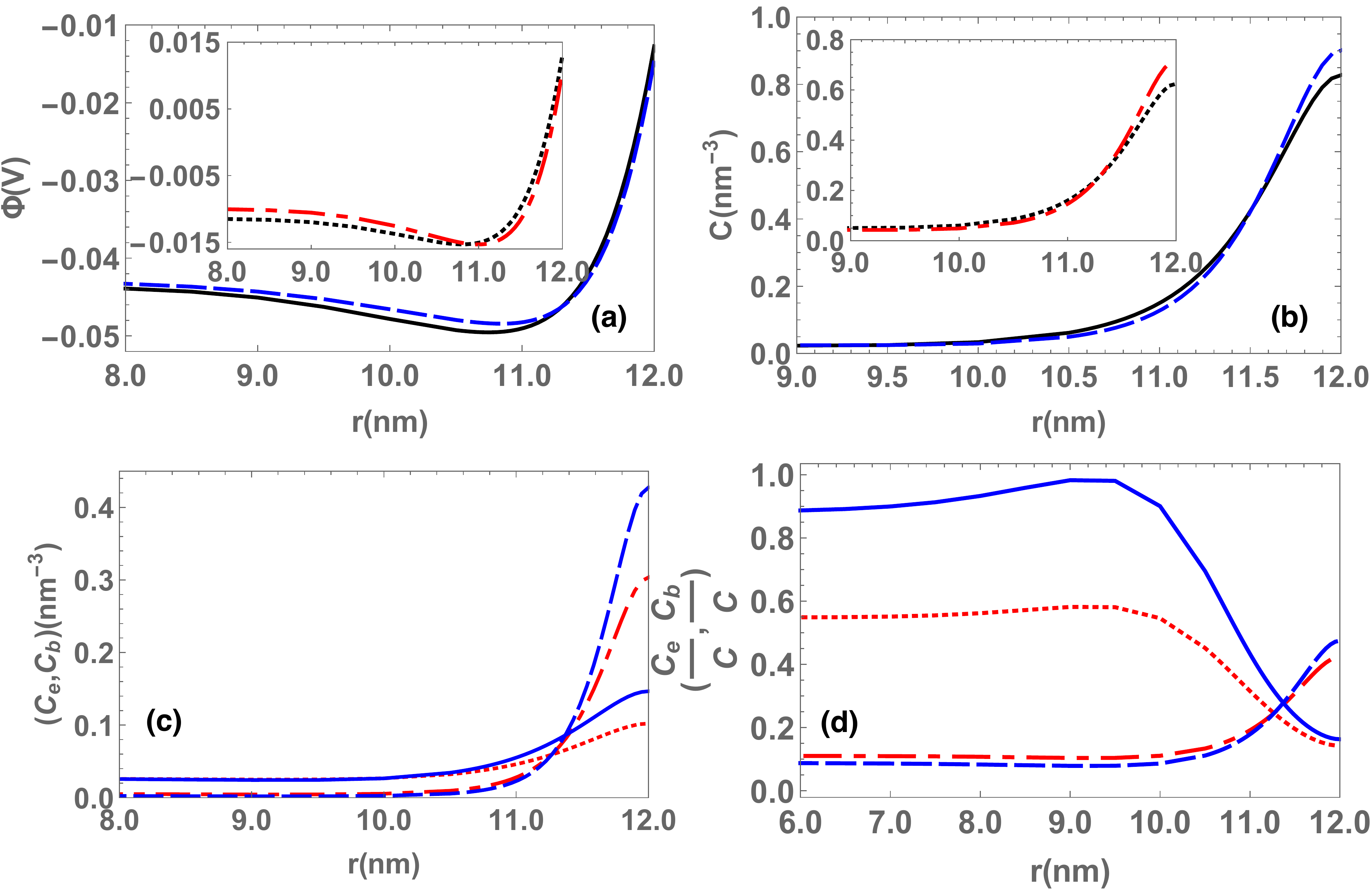}
  \caption{\label{c-phi} For $N=1000$ and two different salt concentrations $\mu$ corresponding to 10 $mM$ (solid and dashed lines) and 100 $mM$ (dotted and dotted-dashed lines),  (a) Electrostatic potential profile for a linear polymer with $f_b=0$ (solid and dotted lines) and branched polymer with $f_b=3.0$ (dashed and dotted-dashed lines) (b) Concentration profile corresponding to two different degree of branching for a linear polymer with $f_b=0$ (solid and dotted lines) and for a branched polymer with $f_b=3.0$ (dashed and dotted-dashed lines).  (c) Concentration profile of endpoints (solid and dotted lines) and branch points (dashed and dotted-dashed lines) for a branched polymer with $f_b=3.0$. (d) Fraction of endpoints (solid and dotted lines) and branch points (dashed and dotted-dashed lines) for a branched polymer with $f_b=3.0$. Other parameters are $\upsilon=0.5$ $nm^3$, $\tau=-1$ $e$, $\sigma=0.4$ $e/nm^2$, $b=12$ $nm$, $a=1$ $nm$ and $T=300$ $K$.}
\end{figure}

We numerically solve the nonlinear coupled differential equations, Eqs.~\eqref{euler_a}, \eqref{euler_b}, \eqref{euler_c}, subject to the constraints given in Eqs.~\eqref{constraint} and \eqref{branch_constraint} to obtain the fields $\Psi$ and $\Phi$ and the parameter $f_e$.   Electrostatic potential and polymer concentration profiles as a function of $r$, the distance from the center of the shell, are shown in Fig.~\ref{c-phi}(a) and (b), respectively for $10 mM$ (solid and dashed lines) and for $100 mM$ (dotted and dotted-dashed lines) salt concentrations for a linear polymer with $f_b=0$ (solid and dotted lines) and a branched polymer with $f_b=3.0$ (dashed and dotted-dashed lines). The total number of monomers enclosed in the shell is $N=1000$ for both profiles shown in the figure. Independent of the amount of salt and degree of branching, the polymer concentration is always larger right next to the surface due to the electrostatic attraction between the polymer and capsid, but it is higher for the branched polymers than the linear one (Fig.~\ref{c-phi} (b)). Note that in all cases the genome profiles remain nearly constant inside the shell but increase noticeably in the vicinity of the capsid wall. 

In addition, we investigated the distribution of branch and end points inside the capsid for $10 mM$ and for $100 mM$ salt concentrations. Figure \ref{c-phi}(c) illustrates the concentration of endpoints $C_e(r)=\frac{1}{\sqrt{a^3}}f_e\Psi(r)$ (solid line for 10 $mM$ and dotted line for 100 $mM$) and branch points $C_b(r) =\frac{\sqrt{a^3}}{6} {f_b}\Psi^3(r)$ (dashed lines for 10 $mM$ and dotted-dashed lines for 100 $mM$),  obtained from Eq.~\eqref{NeNb}. As shown in Fig.~\ref{c-phi}(c), the number of branch points increases in the vicinity of the capsid wall at both salt concentrations; however, it increases even more at the lower salt concentration indicating more segments interact with the wall. The end points, on the other hand, mainly distributed over the interior of the shell. Figure \ref{c-phi}(d) shows the fractions of end points $C_e/C$ (solid lines for 10 $mM$ and dotted line for 100 $mM$) and fraction of branch points $C_b/C$ (dashed lines for 10 $mM$ and dotted-dashed lines for 100 $mM$) as a function of $r$. 

Once the fields $\Psi$ and $\Phi$ are obtained, we insert them into Eq.~\eqref{free_energy} to calculate the free energy of chain-capsid complex, $F$. To obtain the encapsidation free energy, $F$, we need to calculate the free energy of a chain free in solution and that of a positively charged capsid and then subtract them both from the chain-capsid complex free energy, $F$ given in  Eq.~\eqref{free_energy}.  

The capsid self-energy ($F(N=0)$) due to the electrostatic interactions is calculated through Eqs.~\eqref{euler_b} and \eqref{euler_c} in the limit as $N \to 0$, and should be explicitly subtracted from the encapsidation free energy.  The focus of this paper is on the solution conditions in which the capsid proteins can spontaneously self-assemble in the absence of genome as seen in different kind of experiments \cite{Lavelle2009,Zlotnick}. Note that the free energy associated with a free chain (both linear and branched) is negligible under the experimental conditions \cite{Vanderschoot,Siber-nonspecific,Paul:13a}. To avoid the problem of proper free energy rescaling, we furthermore calculate the osmotic pressure of RNA trapped inside the capsid and investigate the impact of its secondary structure on the stability of capsid. Through the calculation of osmotic pressure, we have been able to confirm all our conclusions obtained through the free energy calculation.

\begin{figure}
  \includegraphics[width=8.6cm]{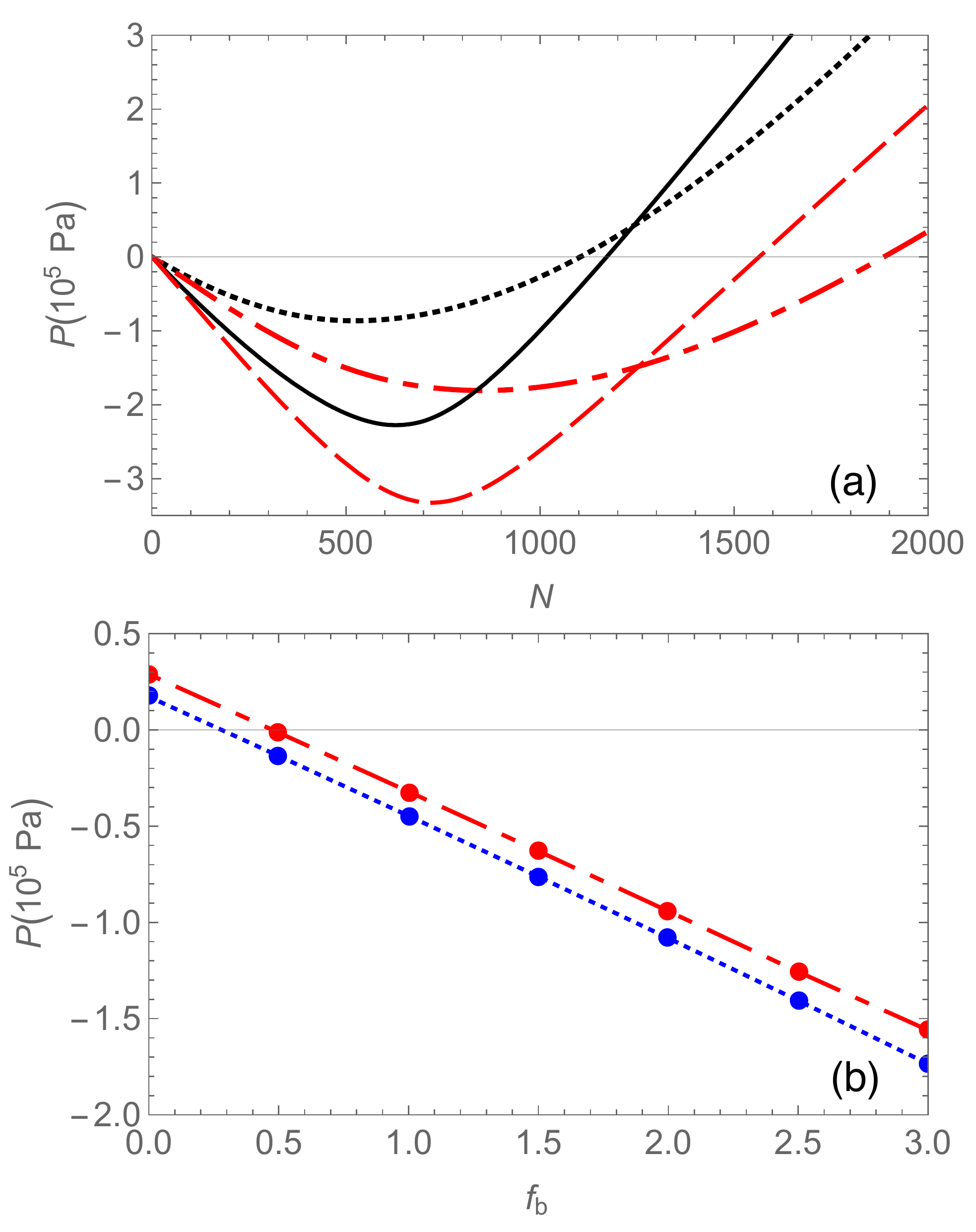}
  \caption{\label{op} (a) Osmotic pressure as a function of monomer numbers for a linear polymer with $f_b = 0$ (solid and dotted lines) and a branched polymer with $f_b = 3$ (dashed and dotted-dashed lines). Solid and dashed lines correspond to the salt concentration, $\mu=10$ $mM$ and dotted and dotted-dashed lines represent the salt concentration, $\mu=100$ $mM$. (b) Osmotic pressure for $N=1200$ as a function of fugacity of branch points, $f_b$, at 10 $mM$  (dotted lines) and 100 $mM$ (dotted-dashed lines) salt concentrations. Other parameters are $\upsilon=0.5$ $nm^3$, $\tau=-1$ $e$, $\sigma=0.4$ $e/nm^2$, $b=12$ $nm$, $a=1$ $nm$ and $T=300$ $K$. }
\end{figure}

In order to get the osmotic pressure, we first calculate the free energy of the system as a function of the monomer number $N$ for both linear and branched chains and then insert it in Eq.~\eqref{eq:op}. A plot of the osmotic pressure $P$ vs.~the monomer number $N$ is given in Fig.~\ref{op}(a) for both linear and branched polymers at two different salt concentrations. The solid and dotted lines correspond to linear polymers with $f_b=0$ and dashed and dotted-dashed lines to branched polymers with $f_b=3.0$. The salt concentrations are 10 $mM$ (solid and dashed lines) and 100 $mM$  (dotted and dotted-dashed lines).  As is clear from the figure, the osmotic pressure goes through a minimum and this minimum is displaced towards longer chains as we increase the degree of branching, {\it i.e.}, more monomers can be encapsidated with increasing $f_b$. For example, the minimum of pressure is at $N\approx523$ for a linear polymer $f_b=0$, and increases to $N\approx851$ for a branched polymer with $f_b=3$ at 100 $mM$ salt. At 10 $mM$ salt, the minimum of the free energy is at $N\approx628$ for $f_b=0$ and at $N\approx719$ for $f_b=3$.  

Figure \ref{op} (b) shows the osmotic pressure in terms of the degree of branching $f_b$ for 10 $mM$ (dotted lines) and 100 $mM$ (dotted-dashed lines) salt concentrations with $N=1200$. When $f_b=0$ (linear polymer), the osmotic pressure is positive but changes the sign as $f_b$ increases regardless of the salt concentration. The figure shows that the pressure becomes more negative as the degree of branching increases indicating that the secondary structure of the genome makes the virus more stable.

\begin{figure}
  \includegraphics[width=8.6cm]{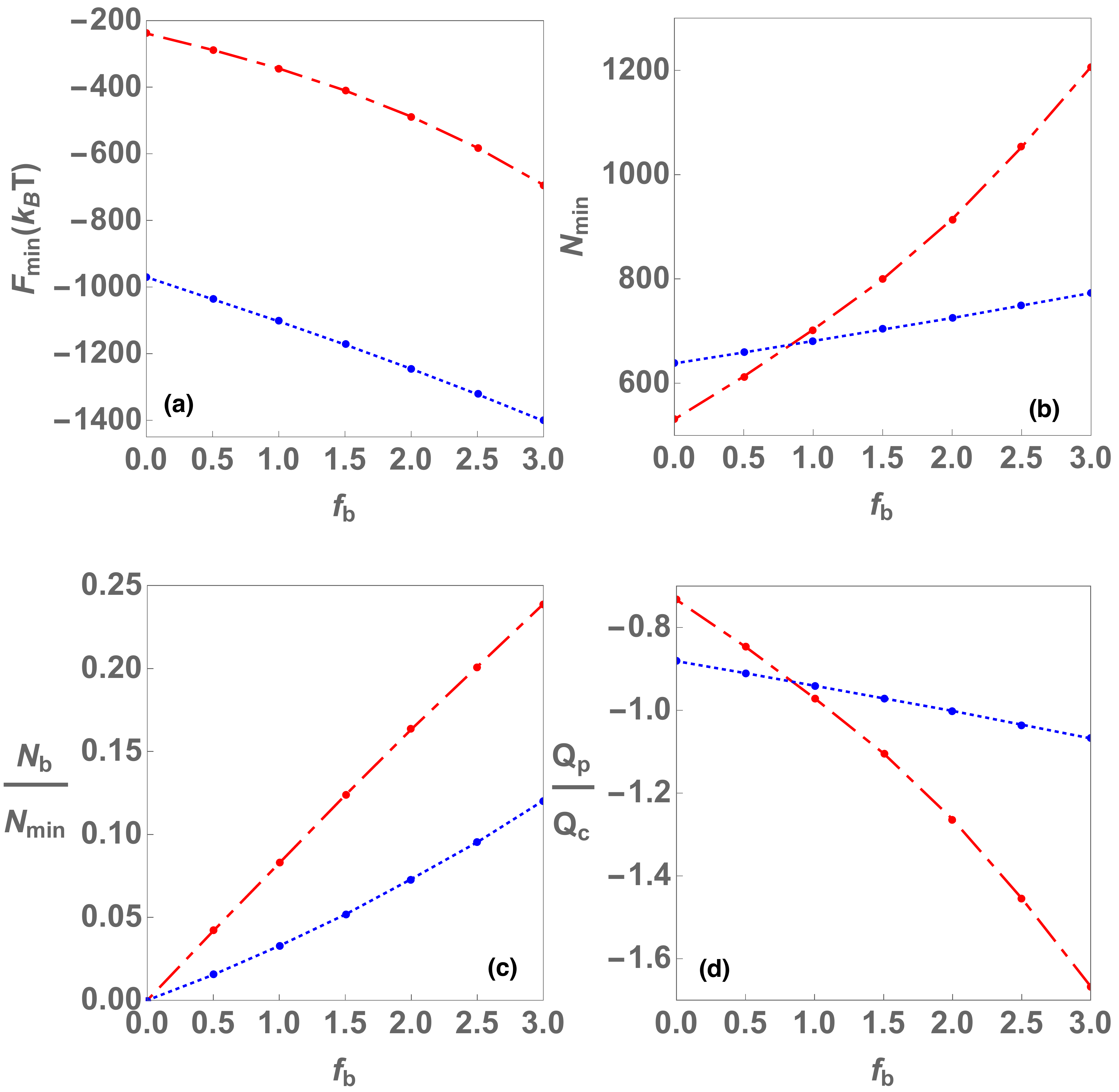}
  \caption{\label{10-100-All} For 10 $mM$ (dotted lines) and 100 $mM$ (dotted-dashed lines) salt concentrations, (a) Optimum free energy (units of $k_BT$) (b) Optimum number of monomers (c) Ratio of number of branched points to the number of monomers at the minima  (d) Ratio of number of polymer charges to the capsid charges at the minima as a function of fugacity of branch points, $f_b$. Other parameters are $\upsilon=0.5$ $nm^3$, $\tau=-1$ $e$, $\sigma=0.4$ $e/nm^2$, $b=12$ $nm$, $a=1$ $nm$ and $T=300$ $K$.}
\end{figure}

To further investigate the role of branching on the assembly of viral shells, we study the impact of branching on the minimum free energy, the optimal number of monomers, the optimal number of branched points, and the ratio of the chain charge to the capsid charge.  A plot of the encapsidation optimum free energy $F_{min}$ vs. the branching fugacity $f_b$ is given in Fig.~\ref{10-100-All}(a) at two different salt concentrations.  For branched polymers, the free energy becomes deeper, indicating that compared to the linear polymers, the branchiness confers more stability to the capsid at both salt concentrations. This effect could explain why some RNAs are encapsidated more efficiently than others, or indeed linear polyelectrolytes. Note that the effect of branching is more apparent at high salt concentrations.  Expectedly, for low salt concentrations, electrostatics overwhelms all the other interactions and the impact of branching becomes less pronounced; nevertheless, the minimum moves towards the longer chains for branched polymers compared to linear ones. 

Figure \ref{10-100-All}(b) shows the optimal number of encapsidated monomers associated with the minimum of free energy as a function of $f_b$. As illustrated in the figure, more monomers are packaged as the degree of branching increases.  For example, at 100 $mM$ for a linear polymer, $f_b=0$, the optimum number of monomers is $N\approx534$ and it increases to $N\approx1211$ for a branched polymer with $f_b=3.0$.  At 10 $mM$ salt, the optimum monomer number for linear polymer is $N \approx 638$ and for branched one is $N_{min} \approx 773$ , with $f_b=3.0$. Figure~\ref{10-100-All}(c) is a plot of the ratio of number of branched points to the optimal number of monomers vs. the branching fugacity. As expected, the ratio increases for higher $f_b$ values. 

The fact that longer, branched chains can be more easily encapsidated by capsid proteins could straightforwardly explain one of the reasons why viruses are overcharged. The total charge of the virion is $Q=Q_p+Q_c= \tau N + 4 \pi b^2 \sigma$ where the first term corresponds to the genome charge and the second one to that of the capsid. Figure~\ref{10-100-All}(d) shows the charge ratio of the genome to the capsid vs. the fugacity of branched points for two different salt concentrations at the minima of the free energy for $\upsilon=0.5$ $nm^3$, $\tau=-1$ $e$, $\sigma=0.4$ $e/nm^2$, $b=12$ $nm$, $a=1$ $nm$ and $T=300$ $K$. The virion becomes overcharged for the values of $f_b>2$ at 10 $mM$ and $f_b>1$ at 100 $mM$. 

%%%%%%%%DISCUSSION%%%%%%%
\section{Discussion and Summary}\label{discussion}
\label{discussion}

We have investigated the role of RNA sequence specificity, as it transpires through the RNA branchiness in the electrostatic encapsidation of RNA viruses. Specifically, we addressed in detail the dependence of the free energy and the osmotic pressure of a confined self-interacting RNA constrained within a spherical, charged  capsid. The sequence specificity was modeled through an annealed distribution of RNA end- and branch-points, and the electrostatics was addressed within a mean-field Poisson-Boltzmann framework,  allowing us to study explicitly the impact of branching and genome-capsid electrostatic interaction on the optimal length of the encapsidated genome. While the details of our model can be subject to criticism and RNA sequence specificity could enter on other more detailed levels of description, we do believe that the coupling between RNA self-interaction and capsid electrostatics represents a robust mechanism of encapsidation and virion stabilization. 

\begin{figure}
  \includegraphics[width=8.6cm]{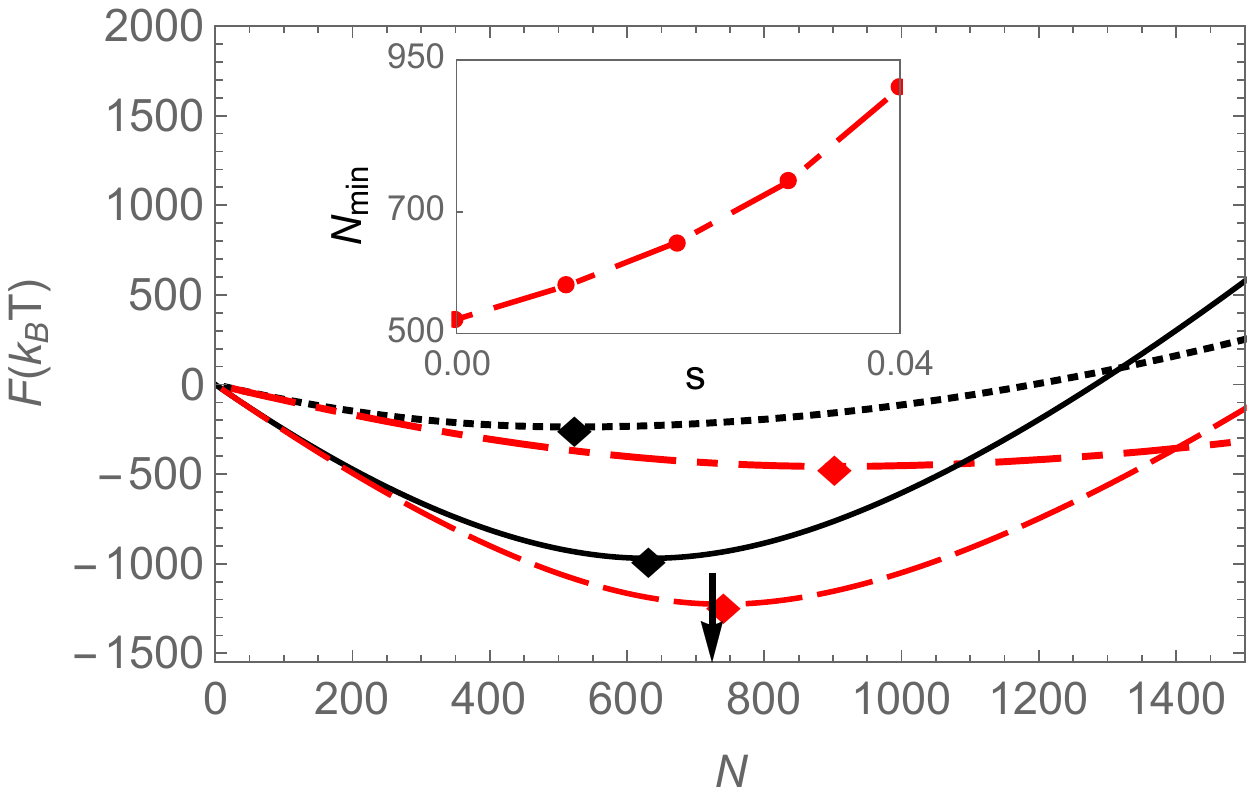}
  \caption{\label{fbasepairing} 
Encapsidation free energy (units of $k_BT$) as a function of monomer number for a self-interacting linear chain model with s=0 (solid and dotted) and s=0.04 (dashed and dotted-dashed lines) at two different values of $\mu$, corresponding to salt concentrations 10 mM (solid and dashed lines) and 100 mM (dotted and dotted-dashed lines). The arrow indicates the monomer number at which the full virus particle is neutral ($Q_p=Q_c$). Inset shows the position of the minimum $N_{min}$ vs. the average fraction of self-paired bases, $s$, for 100 mM salt concentration.  Other parameters are  $\upsilon=0.5$ $nm^3$, $w=1$ $k_B T$, $u=0.5$ $nm^6$, $\tau=-1$ $e$, $\sigma=0.4$ $e/nm^2$, $b=12$ $nm$, $a=1$ $nm$ and $T=300$ $K$.}
\end{figure}

To confirm that the results derived within our model of RNA branching, corresponding to a simple description of the RNA secondary structure, are indeed robust we also propose an alternative self-interacting linear chain model of RNA based on the assumption that RNA can be described as a linear polymer, {\it i.e.}, possesses no branch-points and only two end-points, but self-interacts with short-ranged attractive interactions describing the self-pairing of RNA segments \cite{Borukhov}.  As for the rest, we assume again that the capsid wall can be modeled as a thin, charged spherical shell with uniform surface charge density. The free energy corresponding to this model is again given by Eq.~\eqref{free_energy}, except that the polymer chain is now linear, implying that 
\begin{equation} \label{basepair}
  f_e, f_b \longrightarrow 0,
\end{equation}
and the self interaction term $W[\Psi]$ thus changes to
\begin{equation} \label{basepair}
  W[\Psi] =  \frac{1}{2}(v- a^3 \beta s w) \Psi^4 + \frac{1}{6} u \Psi^6,
\end{equation}
with $s$ the average fraction of self-interacting chain segments, {\it i.e.}, base-pairs, and $w$ is the corresponding short-range binding energy. Note that we included the next, $\Psi^6$ term in the virial expansion in Eq.~\ref{basepair}, with $u > 0$ in order to stabilize the free energy since $(v- a^3 \beta s w)$  can in general  become negative. Variation of the free energy yields the same Euler-Lagrange equations as given in Eqs.~\eqref{euler_a}, \eqref{euler_b}, \eqref{euler_c} subject to the constraint, Eq. \eqref{constraint}. The results of this calculation are presented in Fig.~\ref{fbasepairing} that illustrates the encapsidation free energy as a function of the number of monomers, $N$. As illustrated in the figure, the positions of the free energy minima move towards longer polymers (larger $N$) and the depth of the minima increase with increasing $s$, the average fraction of bound segments. At 10 $mM$ salt, Fig.~\ref{fbasepairing} shows that the minimum of the encapsidation free energy is located at $N=632$ for $s=0$ and at $N=740$ for $s=0.04$.  The effect is again more pronounced at 100 $mM$ salt in which the location of the minimum moves from $N=524$ for $s=0$ to $N=903$ for $s=0.04$. $w$ is chosen $1$ $k_B T$ and $u=0.5$ $nm^6$ in our calculations.

It thus seems that this rather different model, though presenting the same salient features of the system,  yields the same qualitative behavior as discussed above for branched polymers. This substantiates our claim that the coupling between RNA self-interaction and capsid electrostatics represents a robust mechanism of encapsidation and virion stabilization. 

\begin{figure}
  \includegraphics[width=8.6cm]{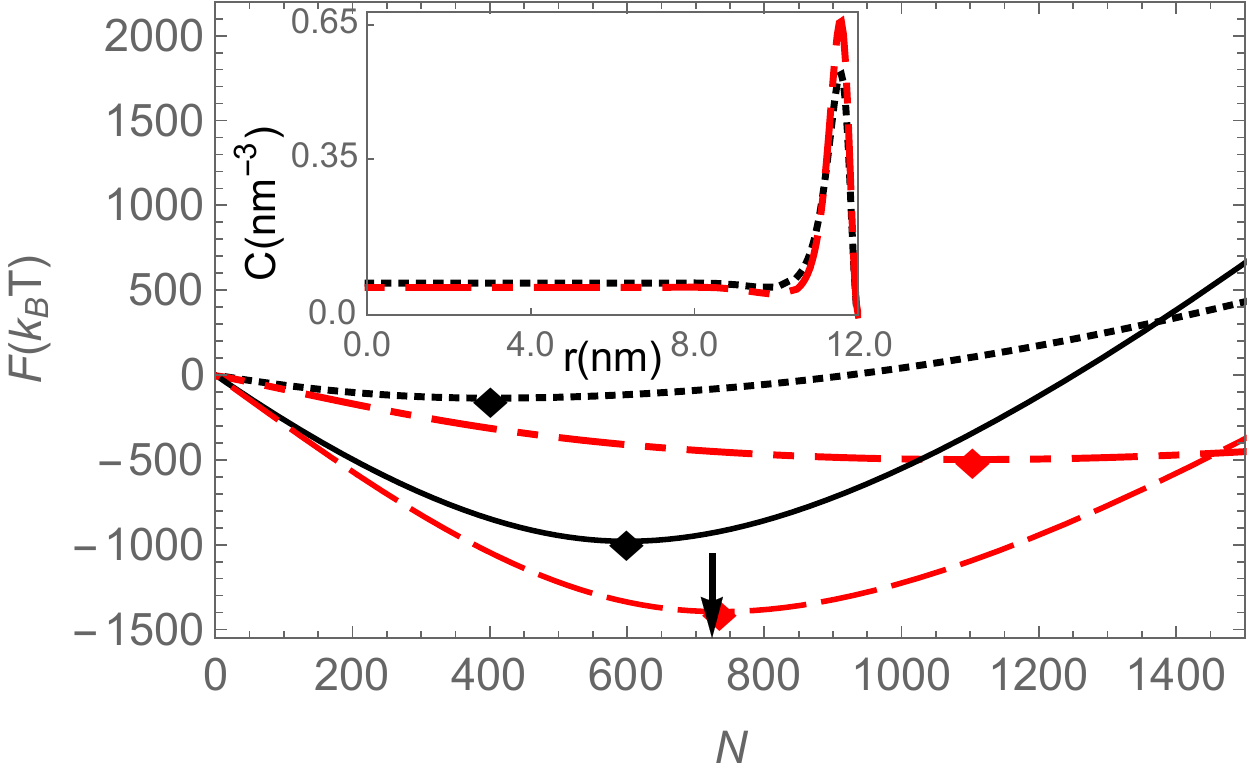}
  \caption{\label{dirichlet} 
Encapsidation free energy (units of $k_BT$) vs monomer numbers for a linear chain with $f_b=0$ (solid and dotted lines) and a branched chain with $f_b=8.5$ (dashed and dotted-dashed lines) at two different sal concentrations $\mu$, 10 mM (solid and dashed lines) and 100 mM (dotted and dotted-dashed lines) with the Dirichlet BC. The arrow indicates the monomer number at which the full virus particle is neutral ($Q_p=Q_c$). Other parameters take the values $\upsilon=0.05$ $nm^3$,  $\tau=-1$ $e$, $\sigma=0.4$ $e/nm^2$, $b=12$ $nm$, $a=0.5$ $nm$ and $T=300$ $K$. Inset shows the concentration profile for N=1000 with two different branching fugacities, $f_b=0$ (linear chain) for the dotted line, and $f_b=8.5$ (branched chain) for the dotted-dashed lines.}
\end{figure}

In addition to investigating the different ways of modeling the secondary structures of RNA, we also studied the impact of different boundary conditions on the encapsidation free energy and osmotic pressure. While all the results presented above correspond to the Neumann BC, ${\hat n.}{\nabla \Psi}|_s=0$, we found that our conclusions do not depend on the type of BCs in that we obtained qualitatively the same results for the Dirichlet BC, $\Psi|_s=0$. Although the Dirichlet BC changes the polymer density profile (see the inset of Fig.~\ref{dirichlet}), the behavior of the free energy and the osmotic pressure remains qualitatively remarkably unaffected in that the minimum of the free energy does get deeper and moves towards longer chains as branching increases.  As is clear from Fig.~\ref{dirichlet}, at 100 $mM$ salt the minimum of the free energy at $N\approx401$ for a linear polymer with $f_b=0$, is displaced to $N\approx1103$ for a branched polymer with $f_b=8.5$ when the Neumann BC is replaced by the Dirichlet BC for the polymer density field. Furthermore, for the Dirichlet BC at 10 $mM$ salt, the free energy minimum is displaced from $N=599$ for $f_b=0$ to $N=735$ for $f_b=8.5$. {Note that the value of $f_b$ used for Dirichlet is chosen such that the ratio of number of branch points to the number of total monomers is almost the same as those for Neumann case.} 

We also calculated the osmotic pressure for Dirichlet BC using both branched and self-interacting linear chains. Consistent with the free energy results, we found that as the degree of branching or the average fraction of self-interacting chain segments increases, the osmotic pressure as a function of $N$ becomes more negative and its minimum moves towards longer chains. 

Further, we examined the impact on the free energy of the capsid surface charge density ($0.3 \leq \sigma \leq 0.9$), polymer charge density ($-2.0 \leq \tau \leq -0.5$) and Kuhn length ($0.5\leq a \leq 2.0$). For both Dirichlet and Neumann BCs, we found that the optimal number of encapsidated monomers for linear chains is always such that number of charges on the polymer is less than those on the capsid, {\it i.e.}, the virus-like particles (VLP) are undercharged. In contrast, we found that the optimal length of the encapsidated branched polymers is larger than that of the linear polymers for all cases examined, resulting in overcharging of VLPs in many cases. We emphasize that while our findings are consistent with previous mean-field PE theories in that the VLPs with a linear polymer is undercharged \cite{Siber-nonspecific}, our results for linear polymers differ from recent numerical simulations \cite{Hagan} and the scaling theories \cite{Paul:13a} on the assembly of viral particles. While the overcharging for linear polymers, observed in Ref. \cite{Paul:13a} is due to the charges on the N-terminals and in Ref.~\cite{Hagan} could be due to the solution conditions or the protein charge distribution, it is found that the branched structure of the polymer enhances overcharging, consistent with our studies. 

It is difficult to determine the topology of large single-stranded viral RNAs in solution, but recent experiments indicate that the secondary structure does play an important role in the efficient packaging of RNA \cite{Comas,Garmann2015}. The secondary structures can be predicted using a number of softwares, such as RNAsubopt (a program in the Vienna RNA package \cite{Vienna}), RNAfold (another program in the Vienna RNA package \cite{Vienna}) and mfold \cite{mfold}. All these software tools, that are progressively unreliable for longer chains, estimate the free energy changes according to the base-pairing and the loop closure of ssRNA and the secondary structure of RNA results from base-pairing of  G, U, C and A nucleotides. RNAfold and mfold calculate the possible sets of base-pairing corresponding to the minimum free energy, while RNAsubopt has an option to generate Boltzmann weighted secondary structures which can be used to calculate a meaningful ensemble average of any quantity. This software was successfully used \cite{Yoffe2008,LucaRudi2015} to calculate the maximum ladder distance (MLD) and we applied RNAsubopt to calculate the thermally averaged number of branch points for RNA1 of BMV and CCMV to shed light on the experiments noted in the introduction on the competition between RNA1 of CCMV and BMV. We generated the ensemble of secondary structures using the RNA1 sequences of both BMV and CCMV obtained form the National Center for Biotechnology Information Genome Database \cite{ncbi}, and then calculated the thermally averaged number of branched points of RNA1 of BMV and CCMV. We found that RNA1 of BMV has 65 branched points vs.~60.5 branched points of RNA1 of CCMV \cite{standard}.  These numbers confirm the experimental results of Comas-Garcia et al. \cite{Comas} that RNA1 of BMV would be preferentially packaged over RNA1 of CCMV.  We note that although these programs were designed for the short RNAs, many important results have been extracted through finding the ensemble average of the desired quantities for viral genomes of length $~2500-10000$ nucleotides \cite{Yoffe2008,LucaRudi2015}.

The theoretical models presented in this paper clearly indicate the important role of the secondary structure of RNA on the assembly of ssRNA viruses. The secondary structure can be indeed invoked to explain the overcharging observed in RNA viruses, while it promotes the efficiency of RNA packaging by increasing the compactness of RNA in order to better fit into a small capsid. As shown above, the secondary structure of RNA clearly effects the osmotic pressure of the capsid; regardless of the details of the model as well as calculational details such as the form of the BCs, we obtain consistently negative osmotic pressures resulting from the presence of the negatively charged chain. The osmotic pressure becomes more negative for a branched polymer compared to the linear one.

Non-specific electrostatic interactions have emerged as the driving force for virus assembly through both the experimental as well as the theoretical studies \cite{Belyi,Garmann2015,Garmann2014,Siber-nonspecific,RNAtopology14}. In our two simple models we generalized the implementation of electrostatic interactions by coupling it to RNA topology. While this is an important step in realism of the modeling, the present level of description still cannot include the specific interactions (or packaging signals) into a complete picture of virus assembly. Further investigations on both specific and non-specific interactions could help understanding the structure of viruses and take steps on the development of antiviral drugs.

\section*{Acknowledgments}
This work was supported by the National Science Foundation through Grant No. DMR-13-10687 (R.Z.). R. P. acknowledges the financial support of the Agency for research and development of Slovenia (ARRS) under Grants No. P1-0055 and J1-7435. The authors would like to thank the Aspen Center for Physics where some of the work was discussed during the Physics and Mathematics of Viral Assembly workshop.

%Appendix

\section{Appendix}

\subsection*{Derivation of the free energy}
We consider RNA as a single polyelectrolyte in a good solvent in the presence of salt ions. There are $N$ monomers of the polyelectrolyte chain, $N^+$ positive and $N^-$ negative salt ions in the solvent. The microscopic degrees of freedom are the position of the monomers (${\bf{r}}(s)$) and positive (${\bf{r_i^+}}$) and negative (${\bf{r_i^-}}$) ions . The partition function can be written as path integral over all configurations
\begin{equation}
\label{partition}
\mathcal{Z}=\int \mathcal{D} {\bf{r}}(s)  \mathcal{D} {\bf{r_i^+}} \mathcal{D} {\bf{r_i^-}} e^{-\beta\mathcal{H}}
\end{equation}
where
\begin{multline}
\label{hamiltonian}
{\mathcal{\beta H}}={\frac{3}{2 a^2}} {\int_0^N {\mathrm{d}} s \, {\bf{\dot r}}^2(s)}
+\: \frac{\upsilon}{2} \int \mathrm{d} {\bf{r}} \, \hat{\rho}^2_m ({\bf{r}})
+\: \int_0^N \mathrm{d} s V({\bf{r}}(s)) \\
+\: {\frac{\beta}{2}} \int\int\mathrm{d} {\bf{r}} \mathrm{d} {\bf{r'}}\ \hat{\rho}_c ({\bf{r}})
\upsilon_c({\bf{r}}-{\bf{r'}})  \hat{\rho}_c ({\bf{r'}}). 
\end{multline}
The first term in Eq.~\eqref{hamiltonian} describes the ideal entropy of the chain, the second corresponds to the short range steric repulsions between monomers and the third term is an external potential acting on the chain. The last term corresponds to the electrostatic interactions between the charges of monomers and ions. In Eq.~\eqref{hamiltonian} $\upsilon_c$ is the Coulomb interaction 
\begin{equation}
\label{upsilon}
 \upsilon_c={\frac{1}{4 \pi \epsilon \epsilon_0}}{\frac{1}{|{\bf{r}}-{\bf{r'}}|}},
\end{equation}
and $\hat{\rho}_c$ is the charge density operator given by
\begin{multline}
\label{charge_density}
\hat{\rho}_c ({\bf{r}}) = \tau \int_0^N \mathrm{d} s \;\; \delta ({\bf{r}}-{\bf{r}}(s)) \\ + e \sum_i^{N^+} \delta ({\bf{r}}-{\bf{r}}_i^+) 
- e \sum_i^{N^-} \delta ({\bf{r}}-{\bf{r}}_i^-)+\rho_0(r).
\end{multline}
Here, $\tau$ is the uniform monomer charge density along the polyelectrolyte and $\rho_0(r)$ is the  charge density of the inner wall capsid in this system. To calculate the following integral in the partition function
\begin{equation}
\label{z2}
\mathcal{Z}_{\text{salt}} = \int {\mathcal{D}} [{\bf{r}}_i^+] {\mathcal{D}} [{\bf{r}}_i^-] e^{-\frac{\beta}{2} \int \int \mathrm{d} {\bf{r}} \mathrm{d} {\bf{r'}}
\hat{\rho}_c ({\bf{r}})\upsilon_c({\bf{r}}-{\bf{r'}})  \hat{\rho}_c ({\bf{r'}})},
\end{equation}
we introduce a local charge density $\rho_c({\bf{r}})$ and its auxiliary field $\phi({\bf{r}})$ using the following identity
\begin{multline}
\label{unity}
 1=\int {\mathcal{D}} [{\rho_c}({\bf{r}})] \delta({\rho_c}({\bf{r}})-{\hat{\rho}}_c ({\bf{r}}))\\
 =\int {\mathcal{D}} [{\rho_c}({\bf{r}})]
{\mathcal{D}} [{\phi}({\bf{r}})] e^{i \beta \int \mathrm{d} {\bf{r}} ({\rho_c}({\bf{r}})-{\hat{\rho}}_c ({\bf{r}})) \phi({\bf{r}})}
\end{multline}
where the second line is the Fourier transform of the delta function. The auxiliary field $\phi({\bf{r}})$ will turn out to be the electrostatic potential. We then replace the density operator $\hat{\rho_c}$ by the corresponding fluctuating density field $\rho_c$ \cite{MoreiraNetz}. Multiplying Eq.~\ref{z2} by Eq.~\ref{unity} and using  Eqs.~\ref{upsilon} and \ref{charge_density} and the Hubbard-Stratonovich transformation, we find
\begin{multline}
\label{z2final}
\mathcal{Z}_{\text{salt}} = \int {\mathcal{D}} [{\phi}({\bf{r}})] \, (\int\mathrm{d} {\bf{r}} e^{-i \beta e \phi({\bf{r}})})^{N^+}
 (\int\mathrm{d} {\bf{r}} e^{i \beta e \phi({\bf{r}})})^{N^-}
\\e^{-\frac{\beta \epsilon \epsilon_0}{2} \int \mathrm{d}{{\bf{r}}} {(\nabla\phi({\bf{r}}))}^2}
e^{-i \beta \tau \int_0^N \mathrm{d}s \phi({\bf{r}}(s)) } e^{-i \beta \int \mathrm{d} {\bf{r}} {\rho_0}(r)\phi({\bf{r}})}.
\\
\end{multline}
We use the same procedure as above to obtain the contribution of excluded volume interaction to the partition function,
\begin{multline}
\label{z3}
e^{-\frac{1}{2} \upsilon \int  \mathrm{d}{\bf{r}} \hat{\rho}_m^2({\bf{r}})}
 \\= \int {\mathcal{D}} [\psi({\bf{r}})] e^{-\frac{1}{2}\upsilon \int \mathrm{d}{\bf{r}} \psi^2({\bf{r}})} e^{-i \upsilon \int_0^N ds \;\psi({\bf{r(s)}})},
\end{multline}
with $\psi$ the auxiliary field representing the monomer density field. Plugging Eqs.~\ref{z2final} and \ref{z3} into Eq.~\ref{partition}, we find the partition function
\begin{multline}
\label{partition2}
 \mathcal{Z}[N^+,N^-]=
 \\ \int {\mathcal{D}} [{\bf{r}}(s)] {\mathcal{D}} [{\phi}({\bf{r}})] {\mathcal{D}} [{\psi}({\bf{r}})]  \, 
 (\int\mathrm{d} {\bf{r}} e^{-i \beta e \phi({\bf{r}})})^{N^+}
 (\int\mathrm{d} {\bf{r}} e^{i \beta e \phi({\bf{r}})})^{N^-}
\\ e^{-\frac{3}{2 a^2} {\int_0^N {\mathrm{d}} s \, {\bf{\dot r}}^2(s)} -\int_0^N \mathrm{d} s V({\bf{r}}(s)) }
\\ e^{-\frac{\beta \epsilon \epsilon _0}{2} \int \mathrm{d}{{\bf{r}}} {(\nabla\phi({\bf{r}}))}^2 -i \beta \tau \int_0^N \mathrm{d}s \phi({\bf{r}}(s)){-i \beta \int \mathrm{d} {\bf{r}} {\rho_0}(r)\phi({\bf{r}})}}
\\ e^{-\frac{1}{2}\upsilon \int \mathrm{d}{{\bf{r}}}   \psi^2({\bf{r}})-i \upsilon \int_0^N ds \;\psi({\bf{r(s)}})}.
\end{multline}
We now switch to the grand-canonical ensemble modifying only the terms associated with the salt ions
\begin{equation}
\label{grand}
 \Xi[\mu]=\sum_{N^\pm}^\infty {\frac{{\mu}^{{N^+}+{N^-}}}{{N^+}!{N^-}!}} \mathcal{Z}[N^+,N^-],
\end{equation}
with $\mu$ the fugacity (density) of the monovalent salt ions related to the concentration of salt ions in the bulk.
Inserting Eq.~\ref{partition2} into Eq.~\ref{grand}, the grand canonical partition function can be written as
\begin{multline}
\label{grand2}
\Xi = \int {\mathcal{D}} [\phi({\bf{r}})] {\mathcal{D}} [{\psi}({\bf{r}})] e^{-\beta{\mathcal{H}_1}
[\phi({\bf{r}}) , \psi({\bf{r}}) ]}\\
\int \mathcal{D}[{\bf{r}(s)}] e^{-\beta{\mathcal{H}_2}[{\bf{r}}(s)]}\\ 
\end{multline}
with the effective free energies
\begin{multline}
\label{chainpart}
\beta{\mathcal{H}_1}[{\bf{r}}(s)]={\int_0^N {\mathrm{d}} s {\Big{(}} \frac{3}{2 a^2}   {\bf{\dot r}}^2(s)}+ V({\bf{r}}(s)) 
  +  i \beta \tau \phi({\bf{r}}(s))
  \\+i \upsilon  \;\psi({\bf{r(s)}}) {\Big{)}}
\end{multline}
and
\begin{multline}
\beta \mathcal{H}_2[\phi({\bf{r}}) , \psi({\bf{r}})]= \int \mathrm{d}{{\bf{r}}}  {\Big{(}}
\frac{\beta \epsilon \epsilon_0}{2}  {(\nabla\phi({\bf{r}}))}^2 + i \beta \rho_0(r)\phi(\bf{r})\\
-2\mu\cos(\beta e \phi({\bf{r}}))+\frac{1}{2}\upsilon\psi^2 {\Big{)}}
 \end{multline}
The polymer part of the partition function is similar to the Feymann integral of the Hamiltonian $\mathcal{H}= -\frac{a^2}{6}\nabla^2+ U({\bf r})$ with the potential $U({\bf r})= V({\bf{r}}) 
  +  i \beta \tau \phi({\bf{r}}) +i \upsilon  \;\psi({\bf{r}}) $ and imaginary time $t \rightarrow is$ \cite{Borukhov}. We assume that the chain is very long (total number of monomers $N\rightarrow \infty$) with a well defined energy gap such that the ground state approximation is valid. Thus, we have  
\begin{multline}
\label{rs_part}
 \int \mathcal{D}[{\bf{r}(s)}] e^{-\beta{\mathcal{H}_1}[{\bf{r}}(s)]} \approx e^{-N E_0}=e^{-N min\{ \frac{<\Psi_0|\mathcal{H}|\Psi_0> }{<\Psi_0|\Psi_0>}\}} \\
 = \exp \bigg{(} - \int \mathrm{d}{{\bf{r}}} \bigg{\{} \frac{a^2}{6} | \nabla   \Psi_0({\bf r}) |^2 + V({\bf{r}})  |  \Psi_0({\bf r}) |^2 \\
 + i \beta \tau \phi({\bf{r}})  |  \Psi_0({\bf r}) |^2 +i \upsilon  \;\psi({\bf{r}} ) |  \Psi_0({\bf r}) |^2  \\-\lambda(\Psi_0({\bf r}) ^2 -\frac{N}{V}) \bigg{\}} \bigg{)} 
\end{multline}
with $\Psi_0$ the eigenfunction and $E_0$ the eigenenergy of the ground state.   The Lagrange multiplier $\lambda$ is introduced to normalize the wave function. Plugging Eq.~\ref{rs_part} into Eq.~\ref{grand2} and integrating out the $\psi$ field, we find the grand canonical partition function as
\begin{equation}
\label{grand3}
\Xi = \int {\mathcal{D}} [\Phi({\bf{r}})]  e^{-\beta{\mathcal{F}}}
\end{equation}
with
\begin{multline} \label{eq:F1}
\beta{\mathcal{F}} = \int \mathrm{d}{{\bf{r}}} \bigg{\{} \frac{a^2}{6} | \nabla   \Psi_0({\bf r}) |^2 + V({\bf{r}})  |  \Psi_0({\bf r}) |^2 
 + \beta \tau \Phi({\bf{r}})  |  \Psi_0({\bf r}) |^2 
 \\+\frac{1}{2} \upsilon  |  \Psi_0({\bf r}) |^4  -\lambda(\Psi_0({\bf r}) ^2 -\frac{N}{V}) 
- \frac{\beta \epsilon \epsilon_0}{2}  {| \nabla\Phi({\bf{r}})} |^2 \\ + \beta \rho_0(r)\Phi({\bf{r}})
-2\mu\cosh(\beta e {\Phi({\bf{r}})}) \bigg{\}} 
\end{multline}
where we introduce the transformation $\Phi\rightarrow i\phi$ with $\Phi$  being the mean electrostatic potential. Due to the absence of an external potential,  $V({\bf{r}})=0$ and the capsid charge density is  $\rho_0({\bf{r}})=\sigma \delta(z)$ with $\sigma$ the surface charge density.  This leads then to Eq.~\ref{free_energy} considering the constraint given in Eq.~\ref{constraint}. Note that Eq.~\eqref{eq:F1} is for a linear chain with $f_1=0$ and $f_3=0$. For branched polymers in the absence of electrostatic interactions, see Ref.~\cite{adsorption2015}.

\bibliography{longer}

\begin{thebibliography}{52}
\bibitem[{sta()}]{standard}
\bibinfo{note}{The standard deviation for the number of branch points in the
  ensemble of RNA secondary structures is around 2.5 for the RNA1 sequence of
  both BMV and CCMV.}
\end{thebibliography}
\end{document}